\documentclass{aa}
\usepackage{graphicx}
\usepackage{epsfig}
\usepackage{natbib}
\usepackage{amssymb}
\usepackage{amsmath}

\def\be{\begin{equation}}
\def\ee{\end{equation}}
\def\lp{\left(}
\def\rp{\right)}
\def\lb{\left[}
\def\rb{\right]}
\def\ov{\overline}

\begin{document}

\title{ The time variation of the fine structure constant: A statistical analysis of
 astronomical data} 

\author{S. J. Landau\thanks{member of  the Carrera del Investigador
    Cient\'{\i}fico y Tecnol\'ogico, CONICET} \and  C. Simeone\thanks{member of  the Carrera del Investigador
    Cient\'{\i}fico y Tecnol\'ogico, CONICET}  } 

\institute{
Departamento de F{\'\i}sica, FCEyN, Universidad de Buenos Aires,
  Ciudad Universitaria - Pab. 1, 1428 Buenos Aires, Argentina\\ 
\email{slandau, csimeone@df.uba.ar}} 

\date{Received; accepted}

\abstract{}{We analyze different data of the variation of the fine structure constant obtained with different methods to check their consistency.}{We test consistency using the modified student test and confidence intervals. We split the data sets into smaller intervals.  A criterion for this selection is proposed.}{Results show consistency for reduced intervals for each pair of data sets considered.}{Results are at variance with the ones obtained considering mean values over the whole interval.}

\authorrunning{Landau \& Simeone }

\titlerunning{A statistical analysis of $\frac{\Delta \alpha}{\alpha}$ data}

\maketitle

\section{Introduction}
\label{Intro}

The attempt to unify all fundamental interactions resulted in the
development of multidimensional theories like  string-motivated field
theories \citep{Wu86,Maeda88,Barr88,DP94,DPV2002a,DPV2002b}, related
brane-world theories \citep{Youm2001a,Youm2001b,branes03a,branes03b},
and (related or not) Kaluza-Klein theories
\citep{Kaluza,Klein,Weinberg83,GT85,OW97}.  Among these theories, there
are some in which the gauge coupling constants may vary over
cosmological timescales. On the other hand, theoretical frameworks based on first
principles, were developed by different authors
\citep{Bekenstein82,Bekenstein2002,CLV01,BSM02,OP02,BM05} to study the variation of the fine structure constant ($\alpha$) or the electron mass ($m_e$).

Different
versions of the theories mentioned above predict different time
behaviors of the fundamental constants. Thus, bounds obtained from
astronomical and geophysical data are an important tool to test the
validity of these theories.

The experimental research can be grouped into astronomical and local
methods. The latter ones include geophysical methods such as the
natural nuclear reactor that operated about $1.8\ 10^9$ years ago in
Oklo, Gabon \citep{DD96,Fujii00}, the analysis of natural
long-lived $\beta$ decayers in geological minerals and meteorites
\citep{Olive04b} and laboratory measurements such as
comparisons of rates between clocks with different atomic numbers
\citep{PTM95,Sortais00,Marion03,Bize03,Fischer04,Peik04}.
The astronomical methods are based mainly on the analysis of high-redshift quasar 
absorption systems. The relative magnitude of the fine splitting of 
resonance lines of alkaline ions is proportional to $\alpha^2 $. \citet{Murphy01b} and \citet{Chand05} have applied
this method to SiIV doublet absorption lines systems at different redshifts. An extension of this method was proposed by \citet{Bahcall04}. These authors use  strong nebular emission lines of O
III to constrain the variation of $\alpha$. Furthermore, this method was improved   by 
comparing transitions of different species with widely far atomic masses 
and led to
 the first results consistent with a time varying
fine structure constant for a range of redshifts ($0.5 < z < 3.5$) \citep{Webb99,Webb01,Murphy03b}. This method is known in the literature as the many multiplet method (MM). 
However,
other recent independent analyses of similar data
\citep{Srianand04,Chand04,Chand06,QRL04} found no variation.
  On
the other hand, the standard MM technique can be revised to avoid the
deficiencies pointed out earlier in the literature \citep{Bahcall04,QRL04}. This
improved method called in the literature as revised many multiplet
(RMM) method, was applied by \citet{QRL04} to a homogeneous sample of FeII lines at redshift $z=1.15$.  Another method,
to test cosmological variation of $\alpha$,  from pairs of Fe II lines observed in individual
exposures from a high-resolution spectrograph was proposed by
\citet{levshakov05} (this method is known in the literature as SIDAM). The authors found no
variation of $\alpha$ at $z=1.84$ and $z=1.15$ \citep{Levshakov06}. However, a recent
reanalysis of spectrum of the quasar Q1101-264 found variability
within $1\sigma$ \citep{Levshakov07}. 
Besides, by comparing  optical and radio redshifts, a bound 
on $\alpha ^2g_p\frac{me}{mp}$ (where $g_p$ is the proton $g$ factor)
can be obtained \citep{CS95,Tzana07}.
Furthermore, comparing molecular and radio lines provides a bound 
on $g_p \alpha^2$ and the most stringent constraints were obtained 
by \citet{Murphy02}. On the other hand, \citet{darling04} reports
bounds on the variation of $\alpha$ at $z=0.2467$ from the satellite
$18$ cm $OH$ conjugate lines. Finally, \citet{kanekar05} compared the
HI and OH main line absorption redshifts of the different components
in the $z=0.765$ absorber and the $z=0.685$ lens toward B0218+357 to
 establish stringent constraints on changes in $F= g_p
\left({\frac{\alpha^2 m_e}{m_p}}\right)^{1.57}$.
Besides, the time variation of the gauge
coupling constants in the early universe can be constrained using data
from the cosmic microwave background (CMB)
\citep{Martins02,Rocha03} and the primordial abundances of
light elements \citep{Ichi02,Nollet}.

In this paper, we would like to focus on the discrepancy between the data on time variation of $\alpha$ from astronomical observations, following a purely statistical criterion. As described above, different methods are able to constrain the variation of $\alpha$. However, to achieve statistical consistency for each fixed redshift interval, the reported value should be the same. The usual assumption is that if there is any time variation of $\alpha$, such variation is constant over all observed redshifts. So, the reported results are means over a range of redshifts. However, theories mentioned above predict different time evolutions of $\alpha$ yielding different variations for different times \citep{Okada,Marciano,Bekenstein82}. Therefore, to test the evolution of $\alpha$ predicted by those kind of theories, not only a mean value over a range of redshifts, but each individual measurement should be considered. However, not all the reported results are consistent. The most important discrepancy in the literature are the results reported by \citet{Murphy03b} and those reported by \citet{Chand04}.   It should be also mentioned that there is an important debate in the literature about the reliability of the \citet{Chand04} results \citep{Murphy07,Srianand07,Murphy08}. The aim of this paper is to test the consistency of different reported values for the variation of $\alpha$ for smaller intervals than the reported ones, in each case. In this way, we will be able to define a reduced interval where consistency can be assured and thus the theoretical prediction for the evolution of $\alpha$ with time within this interval can be tested. The selection of the redshift interval to be tested is not trivial and a method for this selection is also proposed and described in Sects. \ref{tools} and \ref{resultados}.   It is also important to note that observational errors are important and can not be ignored in any statistical analysis.    In Sect. \ref{tools} we describe the statistical tools (which include the observational errors) we use for testing consistency. In Sect. \ref{resultados}, we apply these tools  to check consistency between the data on varying $\alpha$. Not all groups of data can be tested using the modified student test because the requirements of the method for selecting the redshift intervals can not be fulfilled by most groups of data.  Therefore, we compute confidence intervals of a group of data and compare them with a single reported value of another author to check consistency in most of the cases. We find that consistency between pairs of data sets  can be assured for a reduced interval in each case.  In Sect. \ref{discusion}, we discuss our results and write our conclusions.


\section{Statistical tools}
\label{tools}
The question to be answered is whether, for given redshift intervals, two  experiments compared are consistent or not. The appropriate procedure is thus a test for the difference between two population means, which involves a statistic defined in terms of two sample means and two sample variances. However, in some cases, one of the experiments includes, for a given redshift  interval, very little data, and therefore does not allow us to reasonably define a sample mean and a sample variance. In this situation, the procedure to be followed should rather involve a confidence interval constructed from the sample values of the experiment allowing for a statistical treatment.  In what follows, we discuss in detail these two approaches as well as the choice of the sample size (which, consequently, determines the width of the redshift intervals).

\subsection{Student test}
\label{student}

Within a statistical framework, the null hypothesis (two experiments are consistent) can be formulated as
\be
H_0: \mu_1-\mu_2=0
\ee
where $\mu_1,\,\mu_2$
are the (unknown) population means of each experiment for a definite redshift interval.
Because the sample sizes for the available observational data are not expected to be large, and the true variances are not known, to test the hypothesis we must use a $t$ test, which involves the sample variances. Now, the usual $t$ test is not robust to departures from  normality or from equality of variances when the number of values within each sample are not equal. Thus, we  adopt an approximate test starting from the statistic \citep{devore}:
\be
T=\frac{{\ov X_1}-{\ov X_2}}{\sqrt{\dfrac{S_1^2}{m}+\dfrac{S_2^2}{n}}}
\ee
where ${\ov X_1},\,{\ov X_2}$ are the sample mean values for the given redshift interval, $m,n$ are the data numbers of each sample and  $S_1^2,S_2^2$ are, respectively, their sample variances corrected to include the observational errors $e_i$ \citep{Brandt}: 
\be
S^2=\sum p_i\lb e_i^2+\lp x_i-{\ov X}\rp^2\rb,
\ee
\be
p_i^2=\dfrac{\dfrac{1}{e_i^{2}}}{\sum\dfrac{1}{ e_i^{2}}}.
\ee
The rejection region ($RR$) for the two-tailed test  is defined by 
\be
RR: \left\{
\begin{aligned}
T & \leq - t_{\frac{\lambda}{2},\nu}\\
T &  \geq \ \  t_{\frac{\lambda}{2},\nu}
\end{aligned}
\right.\label{RR}
\ee
where the number of degrees of freedom $\nu$ is given by the rounded value  of 
\be
\tilde\nu=\frac{\lp\dfrac{S_1^2}{m}+\dfrac{S_2^2}{n}\rp^2}{\dfrac{(S_1^2/m)^2}{m-1}+\dfrac{(S_2^2/n)^2}{n-1}}
\ee
and $\lambda$ is the (approximate) level of the test \citep{Brownlee}. Thus, $\lambda$ is the approximate probability of Type I error, that is, the probability of rejecting the null hypothesis when it is true. In practice, the algorithm used will yield a level value $\lambda^*$ such that the obtained value of the statistic lies within the associated rejection region (Eq. (\ref{RR}) with $\lambda^*)$. Therefore, at level $\lambda$ the null hypothesis should  be rejected when $\lambda^*\leq\lambda$. 

\subsection{Confidence Intervals}
\label{intervalos}

In some cases, one of the experiments to be tested includes very few data and, therefore, does not allow us to define a sample mean and a sample variance for a given redshift interval. In other cases, the amount of data is statistically very low to consider the results of the modified student test reliable. For  those cases, we introduce a different procedure: Suppose that for a given redshift interval a group of data 1 allows for a statistical treatment, while a group of data 2 does not. To test the consistency of  given observation 2 against observation 1, from the values of group 1 we construct an interval $I$  of confidence $100\,P\%$. Then,
  if the null hypothesis is true, $P=1-\lambda$ is the probability that the result of an observation of group 2  lies within this confidence interval, and  the null hypothesis should  be rejected at level $\lambda$ when this is not the case. The confidence interval is then centered at the mean value ${\ov X}$ of sample 1, and its width is determined by the complement of the rejection region of a two-tailed test. Thus, under the same hypothesis of the preceding subsection, we have  
\be
I=\lp {\ov X}-t_{\frac{\lambda}{2},\,n-1}\dfrac{S}{\sqrt{n}}\,; {\ov X}+t_{\frac{\lambda}{2},\, n-1}\dfrac{S}{\sqrt{n}}\rp
\ee
where $n$ is the number of values of sample 1, and $S^2$ is the sample variance (corrected as above to include the observational errors);  as before, the choice of the $t$ distribution is motivated by the size of the samples, which are not expected to be large.  In practice, we choose  a level  $\lambda$  and the algorithm  yields a confidence interval for this level. Then, we compare the confidence interval obtained from group 1 with each single reported value for the variation of $\alpha$ obtained from group 2.

\subsection{Sample size}
\label{size}

From a statistical point of view, a  possible criterion to estimate the appropriate sample sizes  is to  limit the probability of Type II error, that is, the probability $\beta$ of not rejecting the null hypothesis when it is false. However,  while the probability $\lambda$ of Type I error can be fixed independent of the population or sample values,  the calculation of $\beta$ requires the choice of a definite alternative hypothesis; that is, to determine $\beta$ the inequality $\mu_1\neq \mu_2$ must be specialized as a definite equality $\mu_1-\mu_2=\delta$. Then, once we have precisely determined a definite alternative hypothesis   and chosen a level $\beta$, in some cases we can obtain a simple analytical expression of the required sample size $n$. For a $t$ distribution no simple expression exists, which can easily be understood by recalling that this distribution includes as a parameter the number of degrees of freedom. However if one is only interested in an estimate, an approximate analytical expression is given in terms of a normal distribution; for a two-tailed test for the mean of a population (or for the associated confidence interval) we have (see Ref. \citep{devore}):
\be
n\simeq \left[\left( {\mathrm z}_\frac{\lambda}{2}+{\mathrm z}_\beta\right)\frac {S}{\delta}\right]^2,\label{numero}
\ee     
where $ {\mathrm z}_\frac{\lambda}{2}$, ${\mathrm z}_\beta$ are obtained by inverting a normal N(0,1) distribution (this would slightly underestimate the size $n$, because the $t$ distribution is less peaked than the normal distribution, but this is not relevant if one is not interested in an exact result). In the case that we test  the difference between the means of two populations, if the corresponding two samples are of equal (or at least similar) sizes, formula (\ref{numero}) can also be applied, with  $S=\sqrt{S_1^2+S_2^2}$. In this approach, the choice of the approximate sample size is thus determined by the Type II error that one is to admit for a given departure from the null hypothesis, this departure being measured  by comparison with the sample variance. In practice, for a proposed level $\beta$ we take as a reasonable assumption an alternative hypothesis $\delta\sim S$ and obtain the corresponding sample size $n$; this proves to be consistent with usual choices of what is to be considered as a significant non null result (see the Discussion below).

\section{Results}
\label{resultados}

Table \ref{data} summarizes the bounds for direct measurements of $\frac{\Delta \alpha}{\alpha}$. Table \ref{data2} shows the details of data  that constrain a combination of fundamental constants that include $\alpha$. Unifying schemes predict that the variation of fundamental constants is related, and that their relationship depends on the theoretical framework.  However, we limit ourselves to studying the variation of $\alpha$ and do not consider the possible variation of other fundamental constants. We have excluded the data from quasar emission lines reported by \citet{Bahcall04} and \citet{grupe05} because the individual errors of those data are 2 orders of magnitude above errors of other groups of data. On the other hand, the data reported by \citet{darling04} and \citet{Murphy02} at $z=0.24$ could not be tested either because there are not enough data from other authors at similar redshift to build confidence intervals.

\begin{table*}[!ht]
\renewcommand{\arraystretch}{1.3}
\caption{Groups of data considered  for methods that constrain $\frac{\Delta \alpha}{\alpha}$. }
\label{data}
\begin{center}
\begin{tabular}{ccccc}
\hline
\hline
Method & Redshift or Redshift interval& Number of data  & $\frac{\Delta \alpha}{\alpha} \pm \sigma$&Reference     \\
$(1)$ & $(2)$ & $(3)$& $(4)$ &$(5)$\\ \hline
Many Multiplet & $0.22 < z < 3.6$&   $128$  & $-0.543 \pm 0.116$&$(1)$  \\ 
 Many Multiplet& $0.452 < z < 2.3$&    $23$ &$-0.06 \pm 0.06 $& $(2)$ \\ 
Many Multiplet & $ z=1.15$&   $1$  & $-0.05 \pm 0.24$ & $(3)$  \\  
RMM & $ z=1.15$&   $1$  & $-0.04 \pm 0.46$ & $(4)$ \\ 
Alkali Doublet  & $2 < z < 3$&   $19$  &$-0.50 \pm 1.30$ &$(5)$  \\ 
Alkali Doublet  & $1.9 < z < 2.8$&    $15$& $0.15 \pm 0.43$ & $(6)$  \\ 
SIDAM & $ z=1.15$&   $1$&$-0.007 \pm 0.084$  & $(7)$  \\ 
SIDAM & $ z=1.84$&   $1$  &$0.54 \pm 0.25$& $(8)$  \\ \hline 
\multicolumn{5}{l}{Notes.- Columns: $(1)$ observational method;  $(2)$ redshift or redshift interval;  $(3)$  number of data;  }\\
\multicolumn{5}{l}{ $(4)$ mean value of $\frac{\Delta \alpha}{\alpha}$ and the corresponding $1 \sigma$ error in units of $10^{-5}$;   $(5)$ reference.}\\
 \multicolumn{5}{l}{References: (1) \citet{Murphy03b}; (2) \citet{Chand04}; (3) \citet{Chand06}} \\
 \multicolumn{5}{l}{ (4) \citet{QRL04}; (5) \citet{Murphy01b}; (6) \citet{Chand05}} \\
 \multicolumn{5}{l}{(7) \citet{Levshakov06}; (8) \citet{Levshakov07}} \\
\end{tabular}
\end{center}
\end{table*}

\begin{table*}[!ht]
\renewcommand{\arraystretch}{1.3}
\caption{Groups of data considered for methods that constrain quantities related to $\frac{\Delta \alpha}{\alpha}$.}
\label{data2}
\begin{center}
\begin{tabular}{ccccc}
\hline
\hline
 Method & Redshift or Redshift interval&   Quantity Measured &$\frac{\Delta \alpha}{\alpha} \pm \sigma$ & Reference     \\ 
$(1)$ & $(2)$ & $(3)$& $(4)$ &$(5)$\\ \hline
 Optical and radio lines & $0.23 < z < 2.4$&   $\alpha ^2g_p\frac{me}{mp}$     &$0.32\pm0.50$& $(1)$  \\ 
 Molecular and radio lines & $z=0.69$&   $\alpha ^2g_p$  &$-0.08 \pm 0.27$   & $(2)$  \\ 
 OH conjugate lines and radio lines & $z=0.765$&   $g_p
\left({\frac{\alpha^2 m_e}{m_p}}\right)^{1.57}$ &$0.27\pm0.27$    & $(3)$  \\ 
 OH conjugate lines and radio lines & $z=0.685$&   $g_p
\left({\frac{\alpha^2 m_e}{m_p}}\right)^{1.57}$&$0.11 \pm 0.13$     & $(3)$   \\ \hline
\multicolumn{5}{l}{Notes.- The observational methods consist in comparing the redshift obtained with different spectral lines.} \\
\multicolumn{5}{l} {Columns: $(1)$ observational method specifying which lines are compared; $(2)$ redshift or redshift interval;} \\
\multicolumn{5}{l}{ $(3)$ quantity measured; $(4)$ mean value of $\frac{\Delta \alpha}{\alpha}$ and its corresponding $1 \sigma$ error; $(5)$ reference.}\\
 \multicolumn{5}{l}{References: (1) \citet{Tzana07}; (2) \citet{Murphy02}; (3) \citet{kanekar05}} \\
\end{tabular}
\end{center}
\end{table*}


One of the
major problems of this analysis lies in the selection of the redshift interval to be
tested. The natural choice  is a length equal to the observational error of the measured redshifts. However, at present there are not enough data available to use this criterion. Another possibility would be a bin size  coming from theories that predict time variation of fundamental constants. However, in general, those models have free parameters, which are estimated from observational bounds on varying $\alpha$. Therefore, we decided to use a purely statistical criterion. As explained in Sect. \ref{size}, the amount of data can be determined from limits on Type II error. Now, we fix $\lambda=0.025$, and Eq. \ref{numero} gives $n \simeq 12, 15, 18$ for $\beta=0.1, 0.05, 0.025$ respectively.

\subsection{Student Test}

The selection of redshift intervals  to be tested would proceed as follows: The first interval (to be considered to apply
the test) starts at a redshift $z=a$ with  $b$ width, where $b$ is the minimum length that includes $n$ data ($n$ will depend on the desired value of $\beta$, see above) for both groups of data or for only one group of data in the case of confidence intervals. The following $i$ intervals
will start at redshift $z=a+i*0.1$ and again the value of $b$ is chosen with the same criterion.
For analyzing the results we also define a criterion: i) If all values of $\lambda^*$ are below the desired level for all intervals, we conclude that there is no consistency between both groups of data; ii) if all values of $\lambda^*$ are above the desired level, we conclude that there is consistency between both groups of data and; iii) in case that some values of $\lambda^*$ are below and some values of $\lambda^*$ are above the desired level, we conclude consistency for a reduced interval. We exclude from the consistency interval, all intervals for which $\lambda^* \le \lambda$. We can assure consistency for the remaining interval.

\begin{figure*}
\begin{center}
\epsfig{file=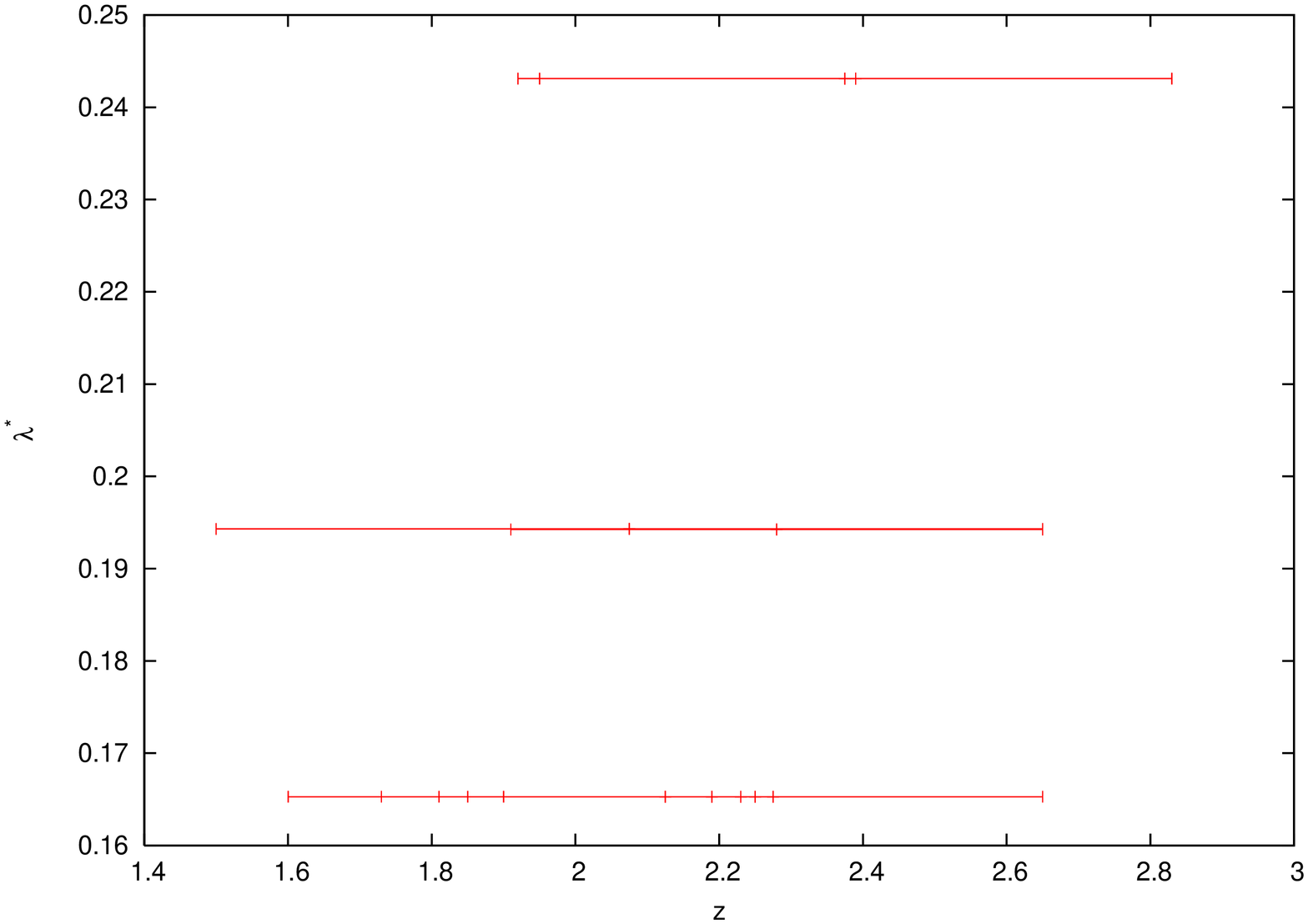,width=2.4in,angle=-90}
\end{center}
\caption{Results of modified student test comparing results from \citet{Chand05} with \citet{Murphy01b}}
\label{grafico2}
\end{figure*}

We applied the modified student test to the data reported by \citet{Chand05} and \citet{Murphy02} for $n=12$  (a higher $n$ is not possible since the total data from \citep{Chand05} are $15$). Results showing consistency for the whole interval are shown in Fig. \ref{grafico2}.  It should be noted that the modified student test can be applied only for redshift intervals larger than $0.76$ due to the requirement that $n=12$. Other groups of data could not be tested in this way due to two kinds of problems: i) the total number of data is lower than $12$; ii) The number of data of one author for a fixed interval is much larger than the number of data of other authors for the same interval. This does not fulfill one of the assumptions to obtain Eq. \ref{numero}.

\subsection{Confidence Intervals}
\label{confianza}

Since the requirements to apply the modified student test are not always fulfilled by the available data, the calculation of confidence intervals is a useful tool to test consistency among data on varying $\alpha$. We  choose  $\lambda=0.025$ and build a confidence interval for a group of data and compare the results with each single reported value of another author. Again, we choose the bin size as the smallest interval centered at the reported value, which contains $n$ data  ($n$ depends on the desired $\beta$ level).  We also define a criterion for analyzing the results. If all confidence intervals  overlap with the reported value of the other author, we conclude that there is consistency for the whole interval, while if none of them overlap, we conclude that there is no consistency. In case there are some confidence intervals that do not overlap with the respective reported interval, we exclude from the consistency interval the corresponding redshift interval.  This is a conservative criterion, because we are probably overestimating the discarded intervals. The available data allow us to perform the analysis for $n=18$ ($\beta=0.025$), and we show these results because the discarded intervals are larger than in the cases for higher values of $\beta$.

Table \ref{table12} shows the confidence intervals calculated for the data of \citet{Murphy03b} for redshift intervals centered in each value of the \citet{Chand04} data and containing $18$ data points (we have also calculated  the confidence intervals for other values of $n$ and results are shown in Sect. \ref{discusion}).   From Table \ref{table12} it follows that the confidence intervals calculated for $z=1.277$, $z=1.541$, and $z=1.637$ do not overlap with the corresponding reported intervals. Therefore, the redshift interval $(1.14,1.93)$ should be discarded from the consistency interval because the data from \citet{Murphy03b} used to calculate the confidence intervals belong to this interval.  Therefore, from the confidence intervals analysis we conclude that there is consistency over the intervals $(0.452, 1.14)$ and $(1.93,2.3)$, i.e., $11$ data points from \citet{Chand04} are consistent with $70$ data points from \citet{Murphy03b} (only $103$ of the $128$ data points are eligible to be tested against \citet{Chand04} data using the criterion $n=18$).

From Table \ref{table7}, it follows that the data reported by \citet{Levshakov07} are at variance with the confidence interval calculated with the data of \citet{Chand04} at $z=1.84$. From Table \ref{table8}, we obtain that the consistency interval for the data sets of \citet{Murphy01b} and \citet{Murphy03b} is $(2.01,2.09)(2.46,2.51)$. From Table \ref{table9}, it follows that the consistency interval for the data sets of \citet{Chand05} and \citet{Murphy03b} is $(1.33,1.71)(2.1,3.11)$. The consistency interval for the data sets of \citet{Murphy01b} and \citet{Chand05} is  $(1.33,1.72)(2.09,3.02)$ (see Table \ref{table10}). Finally, the consistency intervals for the   \citet{Tzana07} and \citet{Murphy03b} data sets calculated from Table \ref{table11} are $(2.27,2.51)$.  These results are at variance with the result  obtained comparing mean values for the whole interval.


\begin{table}[!ht]
\renewcommand{\arraystretch}{1.3}
\caption{Confidence intervals for different redshifts.}
\label{table12}
\begin{center}
\begin{tabular}{cccc}
\hline
\hline
 $z$  &$\Delta z$ & Reported Interval  & Confidence Interval \\ 
 $(1)$ & $(2)$ &$(3)$ &$(4)$ \\ \hline
$  0.452$&$  0.78$&$ (-0.30,0.70)$&$ (-0.96,0.50)$ \\ 
 $  0.822$&$  0.21$&$ (-0.90,0.90)$&$ (-0.78,0.24)$ \\ 
 $  0.859$&$  0.23$&$ (-0.50,-0.10)$&$ (-0.80,0.34)$ \\ 
 $  0.873$&$  0.23$&$ (-0.20,0.20)$&$ (-0.62,0.43)$ \\ 
 $  0.908$&$  0.20$&$ (-0.80,0)$&$ (-0.74,0.25)$ \\ 
 $  0.942$&$  0.18$&$ (-1.90,-0.50)$&$ (-0.88,0.19)$ \\ 
 $  1.182$&$  0.27$&$ (-0.80,0.80)$&$ (-1.51,-0.20)$ \\ 
 $  1.243$&$  0.22$&$ (-0.20,0)$&$ (-1.18,-0.13)$ \\ 
 $  1.277$&$  0.27$&$  (0.,0.20)$&$ (-1.16,-0.20)$ \\ 
 $  1.348$&$  0.35$&$ (-1.00,-0.20)$&$ (-1.17,-0.31)$ \\ 
 $  1.439$&$  0.42$&$ (-0.50,0.50)$&$(-1.14,-0.41)$ \\ 
 $  1.541$&$  0.51$&$ (-0.20,0.20)$&$ (-1.08,-0.40)$ \\ 
 $  1.555$&$  0.48$&$ (-0.30,0.70)$&$ (-0.97,-0.21)$ \\ 
 $  1.636$&$  0.59$&$ (-0.50,0.90)$&$ (-1.21,-0.25)$ \\ 
 $  1.637$&$  0.59$&$  (0., 1.20)$&$(-1.21,-0.25) $ \\ 
 $  1.657$&$  0.60$&$ (-0.20,0.80)$&$(-1.16,-0.14)$ \\ 
 $  1.858$&$  0.49$&$  (0.,0.80)$&$ (-1.38,0.47) $ \\ 
 $  1.915$&$  0.38$&$  (0.50,1.10)$&$ (-1.33,0.76)$ \\ 
 $  2.022$&$  0.44$&$ (-0.50,0.30)$&$ (-1.72,0.69)$ \\ 
 $  2.168$&$  0.41$&$ (-0.40,0.40)$&$ (-1.41,0.78)$ \\ 
 $  2.185$&$  0.38$&$ (-0.10,0.50)$&$ (-1.33,0.91)$ \\ 
 $  2.187$&$  0.38$&$ (-0.40,0.)$&$ (-1.33,0.91)$ \\ 
 $  2.300$&$  0.389$&$ (-0.80,0)$&$ (-1.64,0.50) $ \\ 
\hline
\multicolumn{4}{l}{Notes.- Columns: $(1)$ redshift; $(2)$ length of the redshift } \\
\multicolumn{4}{l}{interval for which the confidence interval is calculated;} \\
\multicolumn{4}{l}{$(3)$ single value reported by \citet{Chand04} }\\
\multicolumn{4}{l}{in units of $10^{-5}$;  $(4)$ calculated confidence interval } \\
\multicolumn{4}{l}{from a group of data reported by \citet{Murphy03b} } \\
\multicolumn{4}{l}{in units of $10^{-5}$.}
\end{tabular}
\end{center}
\end{table}

\begin{table*}[!ht]

\renewcommand{\arraystretch}{1.3}
\caption{Confidence intervals for different redshifts.  }
\label{table7}
\begin{center}
\begin{tabular}{cccccc}
\hline
\hline

 $z$  & $\Delta z$& Reported Interval  & Reference & Confidence Interval & Reference  \\ 
 $(1)$ & $(2)$ &$(3)$ &$(4)$ & $(5)$ & $(6)$ \\ \hline
 $z=1.15$ & $0.24$& $(-0.091,0.077)$& $(1)$ &$(-1.4,-0.06)$ & $(7)$ \\ 
 $z=1.15$ & $1.5$&  $(-0.091,0.077)$& $(1)$&$(-0.13,0.13)$ & $(8)$ \\  
 $z=1.15$ & $0.24$& $(-0.5,0.42)$& $(2)$&$(-1.4,-0.06)$  & $(7)$  \\ 
 $z=1.15$ & $1.4$&  $(-0.5,0.42)$& $(2)$&$(-0.13,0.13)$ &$(8)$   \\  
  $z=1.15$ & $0.24$& $(-0.19,0.29)$&$(3)$ &$(-1.4,-0.06)$ & $(7)$  \\ 
 $z=1.15$ & $1.5$&  $(-0.19,0.29)$&$(3)$ &$(-0.13,0.13)$ &$(8)$   \\ 

 $z=1.84$ & $0.55$&$(0.29,0.79)$ & $(4)$&$(-1.21,0.56)$ & $(7)$  \\ 
 $z=1.84$ & $1.8$& $(0.29,0.79)$ & $(4)$&$(-0.13,0.15)$ & $(8)$  \\ 
 $z=0.69$ & $0.32$& $(-0.19,0.35)$& $(5)$&$(-0.87,0.53)$ & $(7)$  \\  

 $z=0.765$ & $0.26$& $(0,0.54)$& $(6)$&$(-0.79,0.51)$  & $(7)$   \\  

 $z=0.685$ & $0.33$& $(-0.02,0.24)$&  $(6)$&$(-0.99,0.49)$ & $(7)$    \\ \hline
\multicolumn{6}{l}{Notes.- Columns: $(1)$ redshift; $(2)$ redshift interval for which the confidence interval is calculated;   }\\
\multicolumn{6}{l}{$(3)$ single value reported by one author in units of $10^{-5}$;  $(4)$ reference of column $(3)$;}\\ 
\multicolumn{6}{l}{$(5)$ calculated confidence interval from a group of data of another author in units of $10^{-5}$;} \\
\multicolumn{6}{l}{ $(6)$ reference of column $(5)$.}\\
\multicolumn{6}{l}{References: (1) \citet{Levshakov06}; (2) \citet{QRL04}; (3) \citet{Chand06}} \\ 
\multicolumn{6}{l}{(4)\citet{Levshakov07}; (5) \citet{Murphy02}; (6) \citet{kanekar05}} \\
\multicolumn{6}{l}{ (7) \citet{Murphy03b}; (8) \citet{Chand04}} \\ 
\end{tabular}
\end{center}
\end{table*}


\begin{table}[!ht]
\renewcommand{\arraystretch}{1.3}
\caption{Confidence intervals for different redshifts .  }
\label{table8}
\begin{center}
\begin{tabular}{cccc}
\hline
\hline
 $z$   & $\Delta z$ & Reported Interval  & Confidence Interval  \\ 
 $(1)$ & $(2)$ &$(3)$ &$(4)$ \\ \hline
 $  2.01$&$  0.41$&$(-1.04,0.28)$&$(-1.72,0.68)$ \\ 
 $  2.06$&$  0.44$&$(-2.13,1.23)$&$(-1.57,0.81)$ \\ 
 $  2.12$&$  0.40$&$(-1.09,0.39)$&$(-1.43,0.88)$ \\ 
 $  2.14$&$  0.37$&$(-2.56,-0.02)$&$(-1.39,0.82)$ \\ 
 $  2.20$&$  0.40$&$(-1.26,0.08)$&$(-1.33,0.91)$ \\ 
 $  2.28$&$  0.37$&$(0.44,1.64)$&$(-1.56,0.39)$ \\ 
 $  2.30$&$  0.39$&$(-1.03,0.43)$&$(-1.64,0.49)$ \\ 
 $  2.31$&$  0.41$&$(-0.14,1.94)$&$(-1.64,0.50)$ \\ 
 $  2.46$&$  0.33$&$(-0.81,-0.09)$&$(-1.52,0.36)$ \\ 
 $  2.53$&$  0.44$&$(-0.27,1.03)$&$(-1.44,0.38)$ \\
 $  2.62$&$  0.38$&$(-0.90,-0.30)$&$(-1.55,0.23)$ \\ 
 $  2.64$&$  0.40$&$(-1.04,1.08)$&$(-1.14,0.52)$ \\ 
 $  2.77$&$  0.51$&$(0.70,2.10)$&$(-1.63,0.61)$ \\ 
 $  2.81$&$  0.54$&$(-0.44,1.04)$&$(-1.63,0.61)$ \\ 
 $  2.85$&$  0.60$&$(-0.14,0.66)$&$(-1.58,0.67)$ \\ 
 $  2.90$&$  0.69$&$(-2.18, 0.)$&$( -1.48,0.69)$ \\ 
 $  2.91$&$  0.71$&$(-0.67,1.33)$&$(-1.48,0.69)$ \\ 
 $  2.98$&$  0.82$&$(-1.35,2.19)$&$(-1.97,0.59)$ \\ 
 $  3.02$&$  0.84$&$(-1.08,0.28)$&$(-1.97,0.59)$ \\ \hline
\multicolumn{4}{l}{Notes: Column: $(1)$ redshift; $(2)$ length of the redshift interval}\\
\multicolumn{4}{l}{ for which the confidence interval is calculated; $(3)$ single value} \\
\multicolumn{4}{l}{ reported by \citet{Murphy01b} in units of $10^{-5}$; $(4)$  calculated}\\
\multicolumn{4}{l}{ confidence interval from a group of data reported by }\\
\multicolumn{4}{l}{ \citet{Murphy03b} in units of $10^{-5}$.}
\end{tabular}
\end{center}
\end{table}

\begin{table}[!ht]
\renewcommand{\arraystretch}{1.3}
\caption{Confidence intervals for different redshifts. }
\label{table9}
\begin{center}
\begin{tabular}{cccc}
\hline
\hline
 $z$   & $\Delta z$& Reported Interval  & Confidence Interval  \\ 
 $(1)$ & $(2)$ &$(3)$ &$(4)$ \\ \hline
 $ 1.597$&$0.53$&$(-2.92,1.72)$&$(-1.20,-0.32)$ \\ 
 $ 1.908$&$0.39$&$(0.92,15.08)$&$(-1.33,0.76)$ \\ 
 $ 1.915$&$0.38$&$(-18.34,6.74)$&$(-1.33,0.76)$ \\ 
 $ 1.970$&$0.35$&$(-6.74,24.34)$&$(1.61,0.61)$ \\ 
 $ 1.973$&$0.36$&$(-2.64,2.04)$&$( 1.61,0.61)$ \\ 
 $ 1.976$&$0.36$&$(-10.60,10.00)$&$(-1.45,1.05)$ \\ 
 $ 2.168$&$0.41$&$(-2.39,3.39)$&$(-1.41,0.78)$ \\ 
 $ 2.185$&$0.38$&$(-2.60,1.00)$&$(-1.33, 0.91)$ \\ 
 $ 2.329$&$0.44$&$(-13.44,2.24)$&$(-1.92,0.27)$ \\ 
 $ 2.451$&$ 0.35$&$(-0.92,2.92)$&$(-1.46,0.36)$ \\ 
 $ 2.455$&$ 0.34$&$(-6.07,2.07)$&$(-1.52,0.36)$ \\ 
 $  2.456$&$  0.34$&$(-4.83,5.23)$&$(-1.52,0.36)$ \\ 
 $  2.464$&$  0.32$&$(-1.08,3.08)$&$(-1.52,0.36)$ \\
 $  2.493$&$  0.36$&$(-1.09, 10.49)$&$(-1.44,0.38)$ \\ 
 $  2.828$&$  0.58$&$(-5.76,5.16)$&$(-1.63,0.61)$ \\ \hline
\multicolumn{4}{l}{Notes.- Columns: $(1)$ redshift; $(2)$ lenght of the redshift }\\
\multicolumn{4}{l}{interval for which the confidence interval is calculated; } \\ 
\multicolumn{4}{l}{$(3)$ single value reported by \citet{Chand05}}\\
\multicolumn{4}{l}{ in units of $10^{-5}$; $(4)$ calculated confidence interval} \\
\multicolumn{4}{l}{  from a group of data reported by \citet{Murphy03b}} \\
\multicolumn{4}{l}{ in units of $10^{-5}$. }
\end{tabular}
\end{center}
\end{table}

\begin{table}[!ht]
\renewcommand{\arraystretch}{1.3}
\caption{Confidence intervals for different redshifts.  }
\label{table10}
\begin{center}
\begin{tabular}{cccc}
\hline
\hline
 $z$   & $\Delta z$& Reported Interval  & Confidence Interval  \\ 
 $(1)$ & $(2)$ &$(3)$ &$(4)$ \\ \hline
$  1.597$&$  0.53$&$(-2.92,1.72)$&$(-1.20,-0.32)$ \\ 
 $  1.908$&$  0.37$&$(0.92,15.08)$&$(-1.17,0.41)$ \\ 
 $  1.915$&$  0.36$&$(-18.34,6.74)$&$(-1.17,0.41)$ \\ 
 $  1.970$&$  0.34$&$(-6.74,24.34)$&$(-1.31,0.29)$ \\ 
 $  1.973$&$  0.33$&$(-2.64,2.04)$&$(-1.27,0.35)$ \\ 
 $  1.976$&$  0.33$&$(-10.60,10.00)$&$(-1.27,0.35)$ \\ 
 $  2.168$&$  0.26$&$(-2.39,3.39)$&$(-0.74,0.47)$ \\ 
 $  2.185$&$  0.25$&$(-2.60,1.00)$&$(-0.83,0.47)$ \\ 
 $  2.329$&$  0.32$&$(-13.44,2.24)$&$(-0.72,0.34)$ \\ 
 $  2.451$&$  0.28$&$(-0.92,2.92)$&$(-0.85,0.38)$ \\ 
 $  2.455$&$  0.29$&$(-6.07,2.07)$&$(-0.86,0.38)$ \\ 
 $  2.456$&$  0.29$&$(-4.83,5.23)$&$(-0.79,0.36)$ \\ 
 $  2.464$&$  0.31$&$(-1.08,3.08)$&$(-0.79,0.36)$ \\ 
 $  2.493$&$  0.26$&$(-1.09,10.49)$&$(-0.87,-0.04)$ \\ 
 $  2.828$&$  0.38$&$( -5.76,5.16)$&$(-0.50,0.74)$ \\ \hline
\multicolumn{4}{l}{Notes.- Columns: $(1)$ redshift; $(2)$ length of the redshift  }\\
\multicolumn{4}{l}{interval for which the confidence interval is calculated;}\\
\multicolumn{4}{l}{ $(3)$ single value reported by \citet{Chand05}}\\
\multicolumn{4}{l}{  in units of $10^{-5}$; $(4)$ calculated confidence interval}\\
\multicolumn{4}{l}{ from a group of data reported by  \citet{Murphy01b}}\\
\multicolumn{4}{l}{ in units of $10^{-5}$.}
\end{tabular}
\end{center}
\end{table}

\begin{table}[!ht]
\renewcommand{\arraystretch}{1.3}
\caption{Confidence intervals for different redshifts.  }
\label{table11}
\begin{center}
\begin{tabular}{cccc}
\hline
\hline
 $z$   & $\Delta z$& Reported Interval  & Confidence Interval  \\ 
 $(1)$ & $(2)$ &$(3)$ &$(4)$ \\ \hline
  $  0.238$&$  1.21$&$(-0.89,3.31)$&$(-0.96,0.50)$ \\ 
 $  0.312$&$  1.06$&$(-4.87,3.66)$&$(-0.96,0.50)$ \\ 
 $  0.395$&$  0.90$&$(0.07,6.38)$&$(-0.96,0.50)$ \\ 
 $  0.524$&$  0.65$&$(-4.00,-1.90)$&$(-0.96,0.50)$ \\ 
 $  0.525$&$  0.64$&$(-3.41,3.93)$&$(-0.96,0.50)$ \\
 $  1.776$&$  0.58$&$( -3.493,-1.69)$&$(-1.44,0.15)$ \\ 
 $  1.944$&$  0.38$&$(2.85,3.74)$&$(-1.35,0.63)$ \\ 
 $  2.039$&$  0.47$&$(2.43,7.96)$&$(-1.72,0.68)$ \\ 
 $  2.347$&$  0.41$&$(-4.36,-0.72)$&$(-1.70,0.37)$ \\ \hline
\multicolumn{4}{l}{Notes.- Columns: $(1)$ redshift; $(2)$ length of the redshift  }\\ 
\multicolumn{4}{l}{interval for which the confidence interval is calculated;}\\
\multicolumn{4}{l}{ $(3)$ single value reported by \citet{Tzana07}}\\
\multicolumn{4}{l}{  in units of $10^{-5}$; $(4)$ calculated confidence interval}\\
\multicolumn{4}{l}{ from a group of data reported by  \citet{Murphy03b}}\\
\multicolumn{4}{l}{ in units of $10^{-5}$.}

\end{tabular}
\end{center}
\end{table}

\section{Discussion and Conclusion}
\label{discusion}

In Sect. \ref{resultados}  we have obtained two consistency redshift intervals (($0.45, 1.14$) and ($2.03, 2.3$)), when comparing the data sets of \citet{Murphy03b} and \citet{Chand04}.  Now, we apply the modified student test defined in Sect. \ref{student} to the two consistency intervals. We obtain $\lambda^*=0.28$ and $\lambda^*=0.42$ respectively, which confirms the consistency of these two redshift intervals as analyzed using confidence intervals. In this case, we do not have to worry about the number of samples being equal, since the selection of the redshift interval is already done.
It is also interesting to repeat the calculation of confidence intervals for different values of $\beta$ (or $n$). Table \ref{table13} shows that  for a higher value of $\beta$, the consistency interval is larger  than the interval obtained for a lower value of $\beta$.   We also performed the modified student test for the consistency intervals of table \ref{table13} and obtain $\lambda^* > 0.25$.
We have also calculated the consistency intervals for the other groups of data for larger values of $\beta$. We find that there is consistency between the confidence interval and the reported interval for $\beta=0.05$ and $\beta=0.1$ in all cases but the comparison of the data set from \citet{Murphy03b} and \citet{Tzana07}. For this case, the redshift consistency interval for $\beta=0.05$ is $(0.238,1.78)(2.17,2.47)$ and for $\beta=0.1$ we obtain $(0.238,1.79)(2.14,2.47)$. On the other hand, we would like to comment on the comparison between the \citet{Chand05} and \citet{Murphy01b} data sets. In this case, both the modified student test and the calculation of confidence intervals were used to analyze consistency with different results. However, it should be stressed that the modified student test is able to test redshift intervals larger than $0.76$, while the length of the confidence intervals lies between $0.26$ and $0.53$.

A   statistical criterion has been given to estimate the appropriate  sample sizes (which determine the width of the redshift intervals). The criterion is based on the probability of not rejecting the null hypothesis (consistency of two experiments compared) when a definite alternative hypothesis is supposed to be true. While being well justified on statistical grounds,  this proposal  involves a subtle point: The choice of a given alternative hypothesis is not provided by a theoretical framework. Rather, our choice is suggested by the results of \citet{Murphy03b}, which  motivated our analysis. In that work, the non null result for the variation of $\alpha$ was supported by a departure from the null hypothesis  $\delta\sim 4.5 \, \sigma = 4.5 \,S/\sqrt{n}$ ($S$ the sample variance); our choice for the alternative hypothesis $\delta\sim S$ is consistent with this for the employed sample sizes.

We  thus have analyzed the consistency between data sets on varying $\alpha$, using the statistical tools described in Sect. \ref{tools}. Usually, results are taken to be means over a range of redshifts. Here, instead, we have analyzed smaller intervals. The criterion for the selection of intervals is based on limits on the probability of type II error (see section \ref{size}). We have obtained two consistency intervals for the data set of \citet{Murphy03b} and \citet{Chand04}, while the mean values and errors of the whole interval show no consistency.  We have also shown that the data of \citet{Tzana07} and \citet{Murphy03b} are not consistent when tested over small redshift intervals. We have also obtained consistency intervals for this case, which should be used when testing theories on varying $\alpha$. For the other data sets, the consistency depends on the desired value of $\beta$. The result for $\beta=0.025$ is also at variance with the one obtained calculating the mean values and errors over the whole interval. 


\begin{table}[!ht]
\renewcommand{\arraystretch}{1.3}
\caption{Consistency intervals obtained for different values of $\beta$ ($\lambda=0.025$ for all cases).}
\label{table13}
\begin{center}
\begin{tabular}{ccccc}
\hline
\hline
 $\beta$  & $n$ & Consistency Intervals & $n_1$ & $n_2$   \\ \hline
 $0.1$ &$12$& $(0.45,1.17)(1.86,2.30)$  &$12$ &$69$\\ 
 $0.05$ &$15$& $(0.45,1.14) (1.90,2.30)$ &$12$ &$67$\\ 
 $0.025$ &$18$& $(0.45,1.14) (1.92,2.30)$ &$11$ &$67$ \\ \hline
\multicolumn{5}{l}{Notes.- $n_1$ is the number of data reported by  }\\
\multicolumn{5}{l}{\citet{Chand04} that are consistent with $n_2$ data }\\
\multicolumn{5}{l}{from \citet{Murphy03b}.}
\end{tabular}
\end{center}
\end{table}

\section*{{\bf Acknowledgements}}

Support for this work was provided by PIP
5284 CONICET.  The authors would like to thank Amalia Meza for useful discussions about statistics. The authors would also like to thank Michael Murphy for useful discussions about astronomical data.

\newpage

\bibliography{bibliografia}
\bibliographystyle{aa}

\end{document}